\documentclass[11pt]{revtex4}

\usepackage{graphicx}
\usepackage{dcolumn}
\usepackage{bm}

\begin{document}

\title{Possible evidence of non-Fermi liquid behavior from quasi-one-dimensional indium nanowires}

\author{Choongyu Hwang, Namdong Kim, Sunyoung Shin, and Jinwook Chung}

\affiliation{Physics Department, Pohang University of Science and
Technology, San 31 Hyoja Dong, Pohang 790-784, Korea}
\begin{abstract}
We report possible evidence of non-Fermi liquid (NFL) observed at
room temperature from the quasi one-dimensional (1D) indium (In)
nanowires self-assembled on Si(111)-7$\times$7 surface. Using
high-resolution electron-energy-loss spectroscopy, we have measured
energy and width dispersions of a low energy intrasubband plasmon
excitation in the In nanowires. We observe the energy-momentum
dispersion $\omega$(q) in the low $q$ limit exactly as predicted by
both NFL theory and the random-phase-approximation. The unusual
non-analytic width dispersion $\zeta(q) \sim q^{\alpha}$ measured
with an exponent ${\alpha}$=1.40$\pm$0.24, however, is understood
only by the NFL theory. Such an abnormal width dispersion of low
energy excitations may probe the NFL feature of a non-ideal 1D
interacting electron system despite the significantly suppressed
spin-charge separation ($\leq$40 meV).
\end{abstract}


\maketitle
\section{Introduction}
Since the early prediction of non-Fermi liquid (NFL) behavior for an one-dimensional (1D)
interacting electron system,\cite{Voit} numerous efforts have been made to find evidence of NFL
from various forms of 1D conductors including the earlier fractional quantum Hall
system,\cite{chang96} carbon nanotubes,\cite{Ish} self-assembled nanowires on solid
surfaces,\cite{Sego,Lo,Yeom02} and anisotropic bulk materials.\cite{Kim,Aus,Zwick,Hager} The
essential nature of NFL in the 1D electrons systems with enhanced electron correlations is the low
energy bosonic collective excitations rather than the single-particle excitations in Fermi
liquids.\cite{Sam} Photoemission spectroscopy (PES) has been used most extensively to probe
characteristic features of NFL such as the power law behaviors of spectral function near Fermi
level,\cite{Ish,Hager} the spin-charge separation,\cite{Sego,Kim,Aus} and the presence of
pseudo-gap.\cite{Zwick} Since any deviation from an ideal 1D electrons system such as the presence
of impurities, disorder, and thermal fluctuation may sensitively affect the dynamics of these
collective excitations,\cite{Sam,Cap,Hager} continued discussions have been made for some NFL
systems claimed earlier.\cite{Sego,Lo} Although one expects infinite life time for such excitations
in an ideal NFL, added interactions due to any deviations in real 1D systems may cause the damping
of life time $\tau$. Samokhin showed that the damped life time caused by collisions between the
excitations due to the nonlinear band curvature even in a clean NFL system could manifest itself as
the non-analytic dispersion $\zeta(q)$ for the width ($\sim \tau^{-1}$) of a spectral peak stemming
from the excitations.\cite{Sam,Cap} Such a peculiar spectral behavior may sensitively probe the NFL
nature when other spectroscopic evidence is intriguingly suppressed in pragmatic 1D systems.

High energy-resolution ($\leq$ 3 meV) electron-energy-loss spectroscopy (HREELS) has been used to
detect such damped collective excitations, plasmons in particular, in the 1D interacting electron
systems.\cite{Felde,Nagao,Birtsch} In the long wavelength limit, intrasubband plasmon shows an
almost linear energy-momentum dispersion $\omega$(q) as observed experimentally\cite{Goni91} and
also predicted theoretically both by a NFL theory such as a Luttinger theory and by a typical
nearly free-electron gas theory of the random-phase-approximation (RPA).\cite{Sarma96} Despite
numerous theoretical\cite{Sam,Cap,Sarma96,Bor97} and experimental studies\cite{Felde,Goni91,Kuz03}
on plasmon excitations, no unambiguous clue for the NFL behavior due to the damped life time,
however, has been reported to our knowledge.

Here we report evidence of the NFL in the quasi 1D indium (In)
nanowires self-assembled on the Si(111)-7$\times$7 surface at room
temperature. We find a width dispersion $\zeta(q)\sim$~q$^{\alpha}$
with an exponent ${\alpha}$=1.40$\pm$0.24 of the low energy
intrasubband plasmon excitation. This non-analytic width dispersion
is understood only within the framework of NFL theory, never
expected from the RPA.\cite{Sam,Cap,Felde} Unlike the clean NFL,
such a peculiar behavior may arise from the non-ideal features of a
real 1D sample including the quasi-1D nature allowing extra
conduction channels or localization due to impurities and thermal
agitation at finite temperature. It has been, indeed, discussed for
any signature of the NFL from the 1D conducting In nanowires at room
temperature.\cite{Yeom02,Bunk99,Kumpf00,Ahn04,Sak,Cho05}

\begin{figure}[b]
\leavevmode \includegraphics{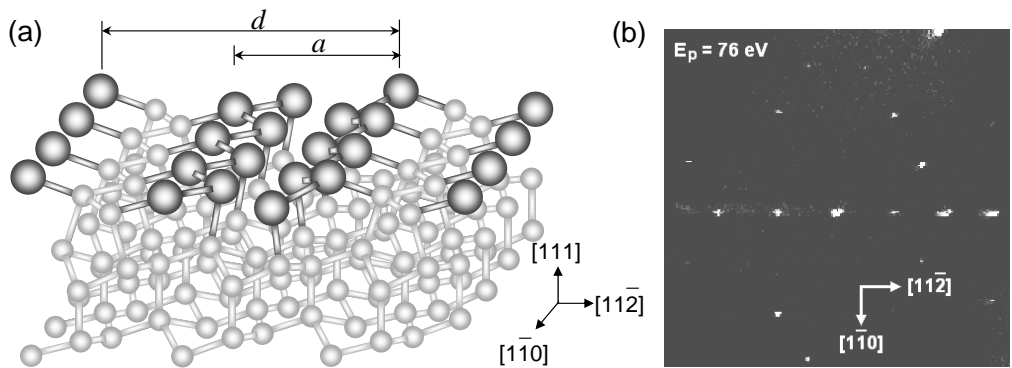} \vspace*{5cm} \caption{(a) Atomic
arrangement of In nanowires of the In/Si(111)-4$\times$1 surface.
Indium atoms forming quasi 1D nanowires are drawn as dark circles
while substrate silicon atoms as grey circles. Four In nanowires
form a bundle of width a=6.9 \AA, and the bundles are separated by
d=13.3 \AA~from each other.\cite{Bunk99} (b) LEED pattern obtained
from the dominant single domain In/Si(111)-4$\times$1 surface.}
\label{fig:structure}
\end{figure}

Although extra interactions due to such deviations may suppress typical NFL features observed in PES studies
significantly, the width dispersion, however, appears to survive and exhibits its unique non-analytic dispersion of the
NFL even at room temperature.

The atomic arrangement of the In nanowires is depicted in
\ref{fig:structure}(a), where four In nanowires form a metallic
bundle along the direction parallel to the nanowires
([1$\bar1$0]).\cite{Bunk99} Note that the In bundles are arranged
periodically along the direction ([11$\bar2$]) perpendicular to the
length of the bundles. Although the transition from the metallic
4$\times$1 at room temperature to the insulating 8$\times$2 phase
below T$_c$=120 K has been understood in terms of Peierls
instabilities accompanying the formation of charge-density-wave
(CDW) with a doubled periodicity,\cite{Yeom02,Cho05,Lee02} some
observations, for example, the partially suppressed spectral
intensity near Fermi level below T$_c$ have not been properly
understood.\cite{Yeom02,Cho05} We ascribe such spectral features to
the intrinsic nature of the NFL phase.

\section{Experiments}
We have obtained our HREELS data by using a Leybold-Heraeus ELS-22
spectrometer under ultra-high vacuum environment of a base pressure
below 1$\times$10$^{-10}$ mbar. The optimum energy resolution and
the half-acceptance angle of the detector are 19 meV and
2$^{\circ}$, respectively. The dispersion data have been obtained by
rotating the sample holder while the monochromator and the analyzer
are fixed in position. Indium atoms were deposited onto the
Si(111)-7$\times$7 surface by thermally evaporating the In rod
wrapped by a tungsten filament. The single domain 4$\times$1 phase
shown in \ref{fig:structure}(b) has been obtained by controlling the
direction of heating current at 400 $^{\circ}$C. By using a high
spatial resolution low energy electron diffraction (SPA-LEED), we
estimate the single domains covering the sample surface more than
93$\%$. No vibrational loss peaks associated with contamination have
been detected during the measurements.

\section{Results and Discussions}
\begin{figure}[h]
\begin{minipage}{18pc}
\includegraphics[width=18pc]{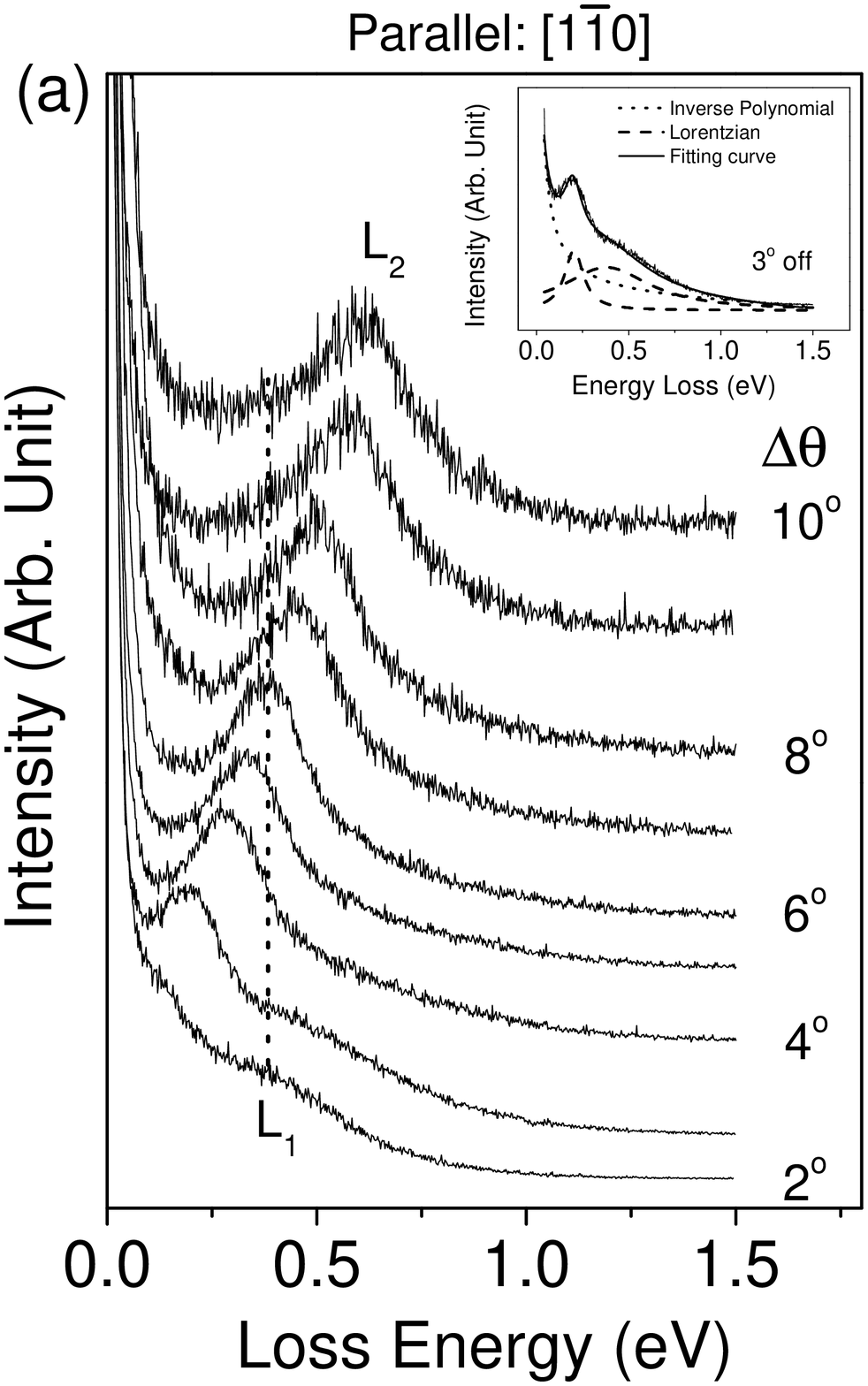}
\end{minipage}\hspace{2pc}%
\begin{minipage}{18pc}
\includegraphics[width=18pc]{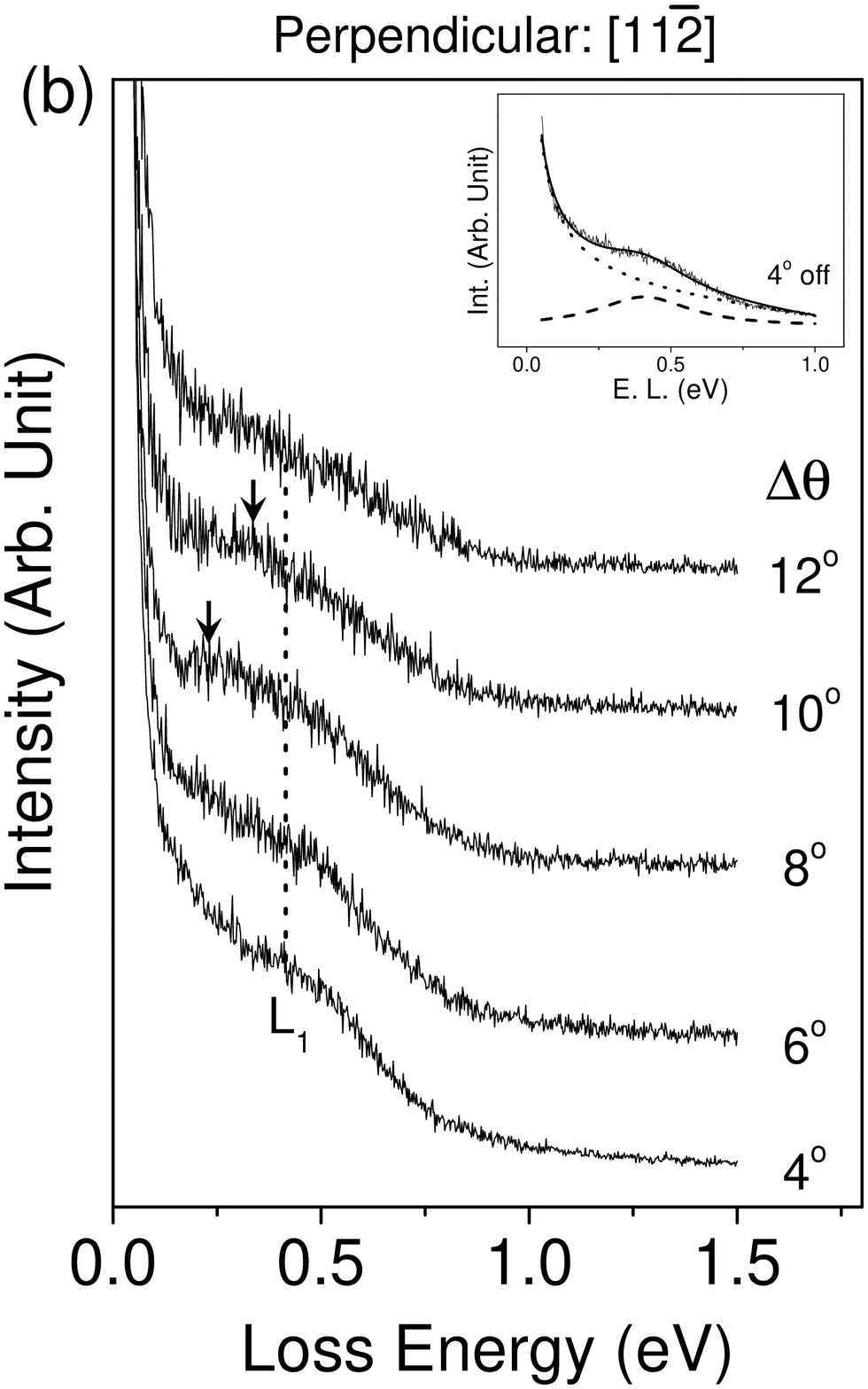}
\end{minipage}
\caption{\label{fig:eels}Angle-resolved EELS spectra obtained from the In/Si(111)-4$\times$1 surface along the parallel
(a) and perpendicular (b) directions to the In nanowires. All spectra are normalized and the primary electron energy is
5.0 eV. $\Delta \theta$ is the angle off from the specular geometry. In (a) two loss peaks of Lorentzian shape, $L_1$
(dashed vertical line) and $L_2$, are found from best fit (see inset) of the spectra while only one non-dispersive
$L_1$ is found in (b). The peaks marked by arrows in (b) are of the same origin with the $L_2$ from minor domains (see
text).}
\end{figure}
We present angle-resolved EEL spectra in \ref{fig:eels} with the
wave vector (q) parallel (a) and perpendicular (b) to the length of
In nanowire. The parallel component of the wave vector
$q_{\parallel}$ is determined by
\begin{eqnarray}
q_{\parallel} = \sqrt{ \frac{2m}{\hbar^{2}} } [(E_{p} - \hbar \omega )^{1/2} \sin \theta_{s} - E_{p}^{1/2} \sin
\theta_{i} ]
\end{eqnarray}
where $E_{p}$ is primary electron energy and $\hbar \omega$ is loss
energy. $\theta_s$ ($\theta_i$) is scattered (incident) angle of the
electron beam. The spectra have been fitted to determine values of
energy and width of the loss peaks as a function of $q$.  Inset in
\ref{fig:eels} shows the Lorentzian fit functions (dashed curves)
and the inverse polynomial background (dotted curves) subtracted
from the raw spectra as done earlier.\cite{Laiten} The resulting
fit-curves (solid curves) are superimposed on the data revealing
excellent fits.

One finds two loss peaks in \ref{fig:eels}(a), a weak non-dispersive
$L_1$ peak of loss energy centered at 378$\pm$11 meV and another
quite dispersive $L_2$ peak along the parallel direction. Note that
there exists only $L_1$ along the direction perpendicular to the In
nanowires. The non-dispersive $L_1$ is ascribed to an interband
transition between the three surface bands $m_1$, $m_2$, and $m_3$
found in photoemission near the Fermi level, because they are nearly
parallel each other with a separation of $\sim$350
meV.\cite{Yeom02,Ahn04} This interband transition then has to be
visible along both directions parallel and perpendicular to the In
nanowires. The significant anisotropic dispersions shown in
\ref{fig:eels} nicely demonstrate the 1D character of the In
nanowires. The loss peaks marked by arrows in the spectra of
8$^{\circ}$ and 10$^{\circ}$ in \ref{fig:eels}(b) are of the L$_2$
origin stemming from the minor rotational domains due to the
120$^{\circ}$ rotational symmetry of a Si(111) surface as discussed
later. The strongly dispersive peak $L_2$ is unique only in the
metallic 4$\times$1 phase since it is absent both in the clean
Si(111)-7$\times$7 and in the In/Si(111)-$\sqrt{31}\times\sqrt{31}$.

One may think of several possible origins for $L_2$; an interband transition, a local atomic
vibration, and a collective excitation such as plasmon or surface phonon. Since the three surface
bands near the Fermi level are quite parallel to each other,\cite{Yeom02,Ahn04} no dispersive
interband transition is allowed for the 4$\times$1 phase. The relatively broad linewidth of $L_{2}$
($\geq$137 meV) rules out the local vibrational origin of width typically less than 25 meV. Since
the loss energy of $L_2$ (up to 619$\pm$23 meV) far exceeds the highest energy of optical phonon of
Si crystal ($\sim$ 60 meV) in addition to the absence of multiple phonon peaks, one may safely
consider $L_2$ as a plasmon due to the collective excitation of conduction electrons along the
nanowires. Furthermore this peak should not be associated with the well known Peierls instability
of a 1D metallic system since the range of momentum measured (0 \AA$^{-1}\leq$q$\leq$ 0.034
\AA$^{-1}$) is far from the wave vectors $2k_F$=0.82 \AA$^{-1}$ for the intrinsic Peierls
instability\cite{Yeom02} or that of the Landau damping where plasmon can decay directly into single
particle-hole excitations as will be discussed later.\cite{Felde,Nagao}. Therefore the plasmon
observed is not affected by such singularities.\cite{Sarma96}

We then compare our experimental energy dispersion curves $\omega(q)$ of the plasmon to the one
predicted by the NFL theory, which is given below for a 1D interacting electron system near Fermi
level.\cite{Voit}

\begin{eqnarray}
\omega(q) = q [v_{F}^2 + \frac{2}{\pi \hbar} v_{F} V(q)]^{1/2}, \label{plasmon}
\end{eqnarray}
where
\begin{eqnarray}
V(q) = \frac{2e^{2}}{4\pi \epsilon} K_{0}(qa).
\end{eqnarray}

$v_{F}$=$\hbar k_F / m^*$ is the 1D Fermi velocity with effective
mass m$^*$, $V(q)$ is the Coulomb potential, and $K_0$ is the
modified Bessel function of the second kind. Note that the spin
excitation alone gives only the first term in \ref{plasmon} so that
the spin-charge separation is due essentially to the second term
when V(q)$\neq$0. Incidentally one finds that the RPA for a 1D
nearly free-electron gas gives the same q-dependence as in
\ref{plasmon} in the small q limit.\cite{Sarma96} Therefore the
energy dispersion $\omega(q)$ alone would not distinguish the NFL
from the RPA.

\begin{figure}[h]
\begin{minipage}{18pc}
\includegraphics[width=18pc]{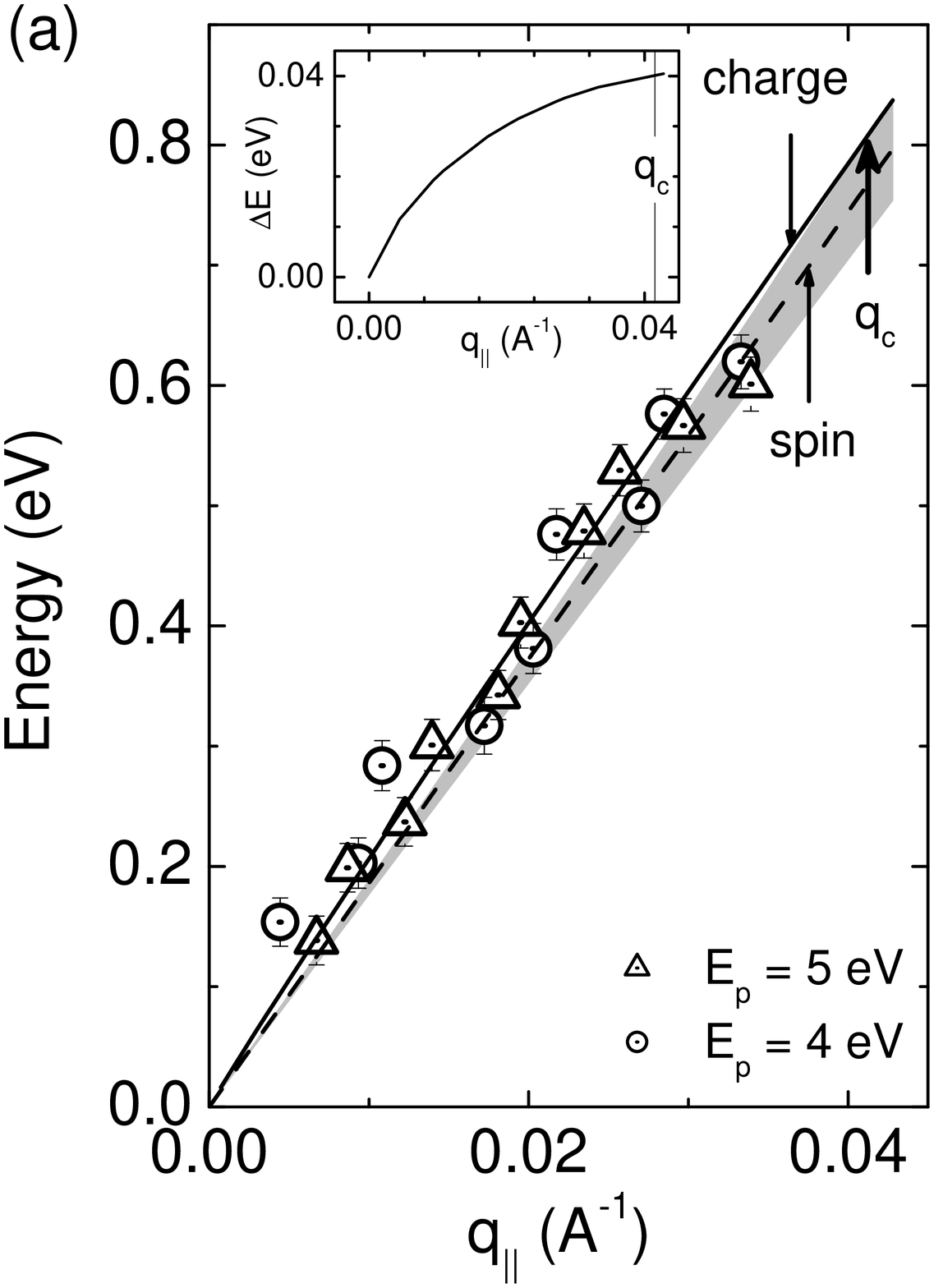}
\end{minipage}\hspace{2pc}%
\begin{minipage}{18pc}
\includegraphics[width=18pc]{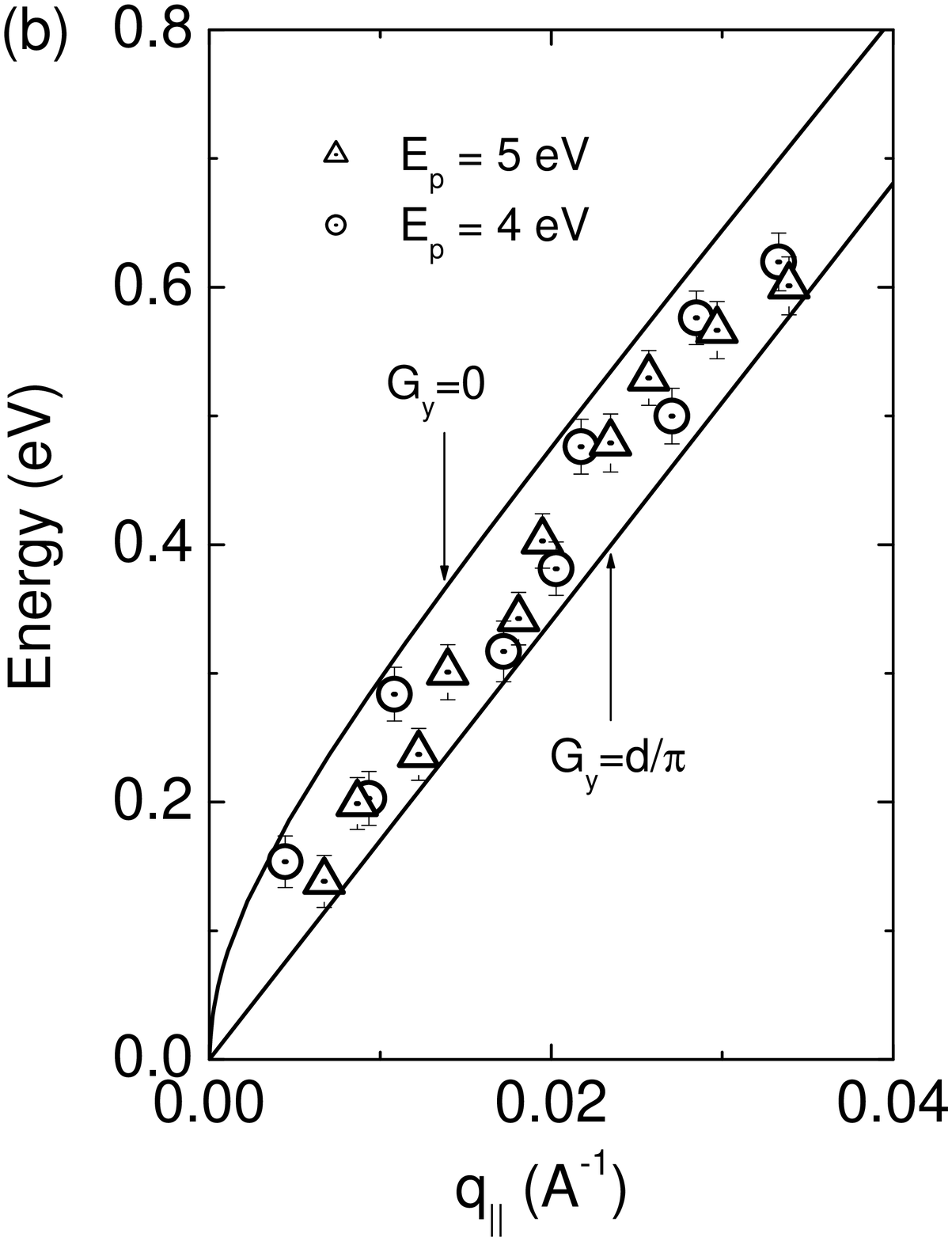}
\end{minipage}
\caption{\label{fig:fit}Fitting the loss energy versus momentum data
(empty circles and triangles) with a theoretical formula $\omega(q)$
in \ref{plasmon} without (a) and with (b) inclusion of inter-In
bundle interactions. In (a), the theory accounts for our data quite
well although the spin (dashed) and charge (solid) separation
$\Delta E$ is relatively small ($\leq$40 meV, see inset). Shaded
area corresponds to the particle-hole excitation continuum. In (b),
all the data points for $q \leq q_c$ are found within the two
limiting dispersions for $G_y$=0 and $\pi/d$ when inter-bundle
interactions are included. }
\end{figure}

Using $a$=6.9 \AA~, we have fitted our energy dispersion data with
\ref{plasmon} floating $m^*$ as a fitting parameter. The results
depicted in \ref{fig:fit}(a) exhibits charge (solid) and spin
(dashed) dispersions with a maximum spin-charge separation of about
40 meV (see inset). One also notes that all our data points locate
below the estimated Landau edge $q_c$=0.041 \AA$^{-1}$ determined by
the crossing q-value between the charge dispersion and the upper
edge of the single particle-hole excitation.\cite{Sarma96} Naturally
the two dispersions merge to $\omega$=0 as q approaches 0. The spin
dispersion found within the single particle-hole excitation and the
relatively small spin-charge separation ($\Delta E\leq$40 meV) may
explain why previous photoemission study with energy resolution
greater than 100 meV has failed to detect any spin-charge separation
at room temperature.

Despite the hidden evidence of NFL in previous photoemission study, \cite{Yeom02} some observations still challenge a
possibility for the NFL behavior of this 1D metal system. Unlike the bands $m_2$ and $m_3$ showing a typical Fermi
liquid behavior, the band $m_1$, despite its typical Fermi edge at room temperature, becomes severely quenched in
spectral intensity at low temperature similar to the NFL behavior of the carbon nanotubes.\cite{Ish} Moreover, the
asymmetric parameter 0.09 obtained for $m_1$ in reference \cite{Yeom02} remains unaltered despite the phase transition
upon cooling, which is not properly understood in terms of either CDW mechanism or metal-insulator transition of Fermi
liquids. Previous EELS study challenges no CDW gap or a possibility of partly metallic phase at low
temperature.\cite{Sak} Another theory paper even predicts a metallic surface due to $m_1$ band for both above and below
$T_c$ partially supporting the idea suggested by the EELS study.\cite{Cho05} Such a so-called pseudo-gap feature
suggested by the significantly depressed spectral intensity near the Fermi level of an interacting 1D electrons system
may be another clue for NFL as for the TTF-TCNQ (tetrathiafulvalene-tetracyanoquinodimethane).\cite{Zwick}

Now the effective masses $m^*$ obtained from the best fits for each
of the three parallel surface bands $m_1$, $m_2$, and $m_3$ are
$m^*/m \sim$ 0.025, 0.11, and 0.17, where $m$ is the mass of a free
electron. The Fermi wave vectors k$_F$ associated with these are
found to be 0.75, 0.54, and 0.41 \AA$^{-1}$,
respectively.\cite{Ahn04} The 1D nature of the $L_2$ is seen also
from the two peaks marked by arrows in \ref{fig:eels}(b) originated
from minor domains rotated by $\pm$ 60$^{\circ}$ from [1$\bar1$0]
direction. When the data points for these loss peaks are projected
onto the wire direction of the single domain by multiplying
cos(60$^{\circ}$) to $q$, they appear to fall onto the same
dispersion curve of the $L_2$ in \ref{fig:fit}(a).

We now consider a so-called ``band effect" by including the
inter-bundle interactions along the [11$\bar2$] direction. With a
separation d=13.3 \AA~ between neighboring bundles, we have modified
$V(q)$ by adding the second term in \ref{bandeffect} in the long
wavelength limit.\cite{Sarma96}

\begin{eqnarray}
V(q) & = & \frac{2e^{2}}{4\pi \epsilon} \{K_{0}(qa) + 2\sum_{s=1}^{\infty}K_{0}(sqd)\cos(G_y sd)\} \label{bandeffect}
\end{eqnarray}

where $G_y$ is the lattice vector perpendicular to the nanowires. As
seen in \ref{fig:fit}(b), we find all the data points for $q \leq
q_c$ are contained within the two limiting dispersion curves for two
values $G_y$=$0$ and $\pi/d$. We thus conclude that the loss peak
$L_2$ is an intrasubband plasmon excitation with its energy
dispersion well described either by the RPA or the NFL theory for $q
\leq q_c$.

\begin{figure}[h]
\begin{minipage}{18pc}
\includegraphics[width=18pc]{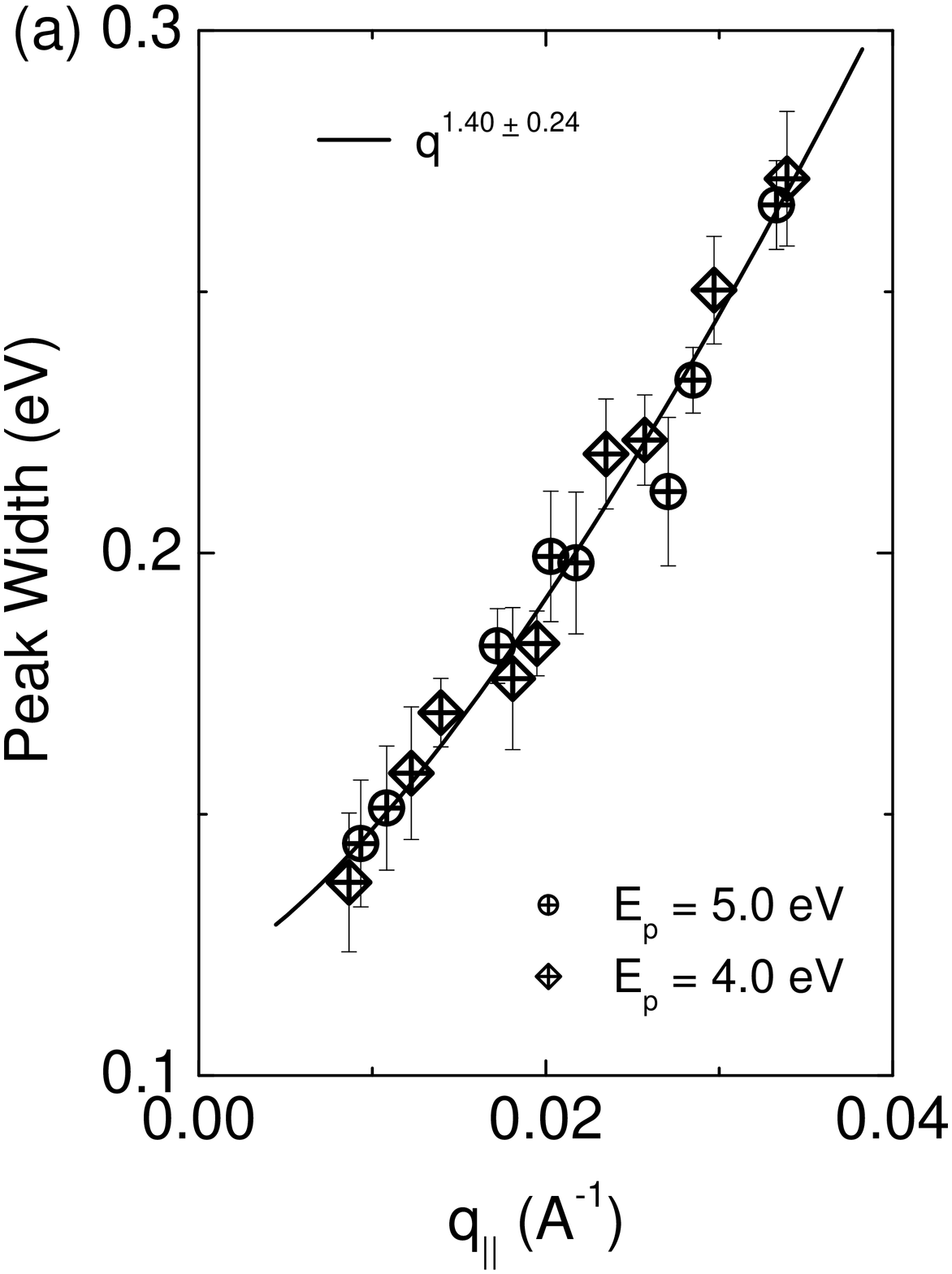}
\end{minipage}\hspace{2pc}%
\begin{minipage}{18pc}
\includegraphics[width=18pc]{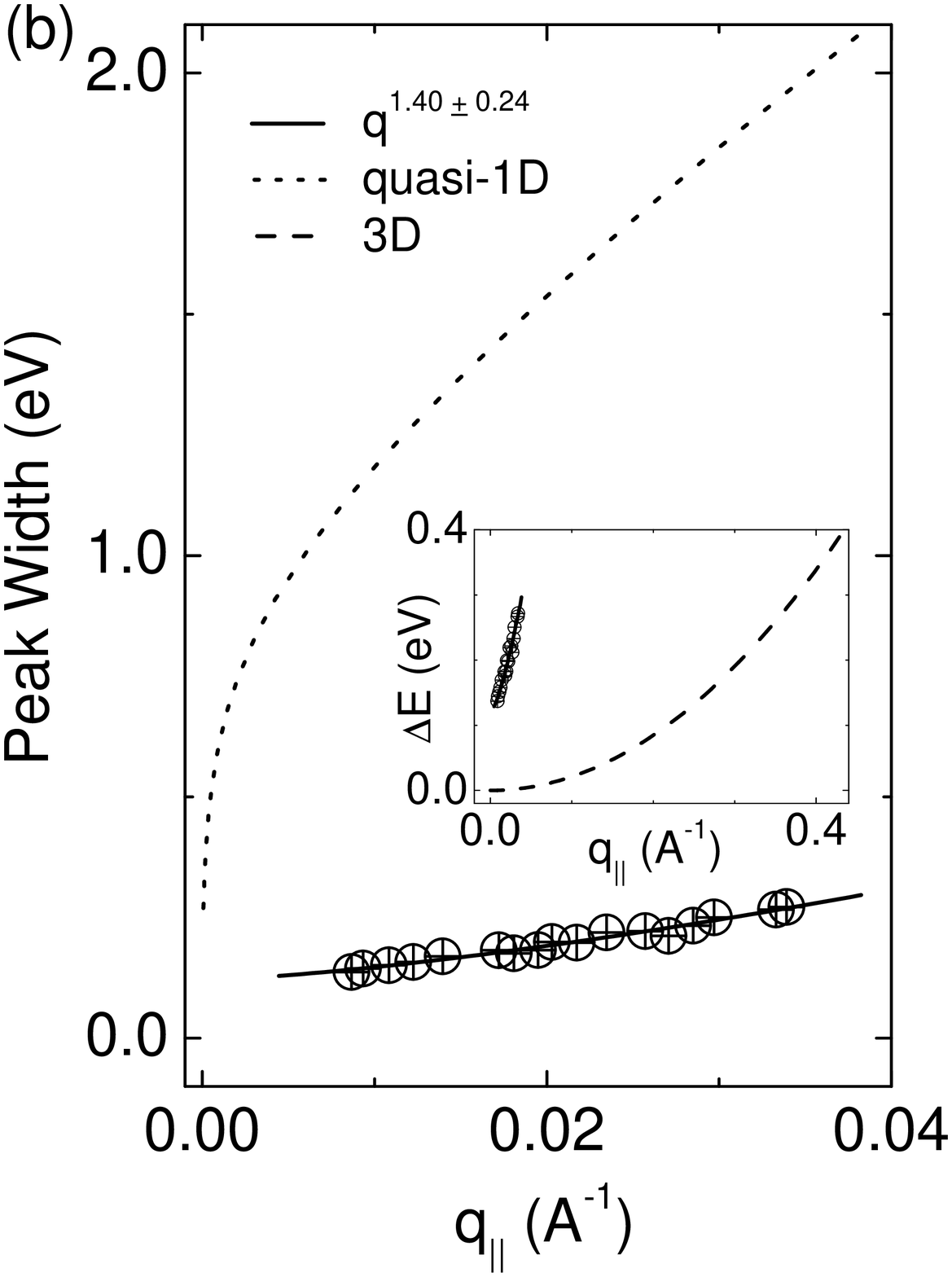}
\end{minipage}
\caption{\label{fig:width}(a) Results of the fits of our width data (crossed squares and circles) of the intrasubband
plasmon with a theoretical fit function $\zeta(q)\sim q^{\alpha}$. The best fit curve (solid curve) gives
${\alpha}$=1.40$\pm$0.24. (b) The estimated dispersions for nearly free-electron gas in quasi-1D (dotted curve) and 3D
(dashed curve in inset) electron systems.}
\end{figure}

Although the energy dispersion $\omega(q)$ shows identical behavior
for both NFL and nearly free-electron gas in the low q limit, the
width dispersion $\zeta (q)\sim$ q$^{\alpha}$ of the NFL, however,
clearly distinguish the NFL nature from nearly free-electron gas
when the exponent ${\alpha}$ appears to be
non-analytic.\cite{Sam,Cap} In order to determine this crucial
element $\alpha$, we have fitted two sets of width values of the
$L_2$ with $\zeta (q)\sim q^{\alpha}$. The best fit (solid curve)
gives ${\alpha}$=1.40$\pm$0.24 as shown in \ref{fig:width}(a). As
discussed below, this non-analytic width dispersion, in fact,
strongly supports the NFL description of the 1D conducting In
nanowires at room temperature.

The width of a plasmon excitation may be changed by various possible
causes such as disorder, instrumental resolution, and superposition
of multiple bands. However, we have excluded these possibilities,
since the surface phase remained unchanged maintaining the single
domain ($\geq$ 93 \%) ordered In/Si(111)$-$4$\times$1 phase during
the entire course of the measurements done with a fixed instrumental
resolution ($\Delta q = 0.0038 \pm 0.0002$ \AA$^{-1}$). This was
confirmed by observing both no noticeable spectral changes at a
fixed momentum and no loss peaks associated with contamination or
local defects before and after measurements. The quite symmetric
line-shape of the $L_2$ peak for all $q$ values obtained by
subtracting the $L_1$ as well as the background intensity from the
raw spectra excludes also the possibility of broadening due to the
superposition of multiple bands.

We now discuss physical significance of a non-analytic exponent
${\alpha}$ in real 1D conducting samples. It has been predicted
${\alpha}$=1.5 for a NFL system with enhanced collisions between
bosonic excitations due to non-linear band curvature, which is quite
distinct from the clean NFL.\cite{Voit,Sam,Felde} This may be quite
plausible for the In nanowires since the three surface bands $m_1$,
$m_2$, and $m_3$ of the In/Si(111)-4$\times$1 surface have quite
non-linear band dispersions. We note also that our width dispersion
with ${\alpha}$=1.40$\pm$0.24 is clearly distinguished from the
non-linear dispersion of quasi-1D nearly free-electron systems where
two particle-hole pairs excitations play a vital role.\cite{Tan}
Such a non-linear dispersion is drawn in \ref{fig:width}(b) by the
dotted curve, which apparently is far from our experimental data.
Obviously the dispersion with ${\alpha}$=2 for volume (3D) plasmons
in conventional metals (see inset in \ref{fig:width}(b)) may also be
safely ruled out since it deviates too far from experimental
data.\cite{Sturm} Notice that our width dispersion in
\ref{fig:width}(a) seems to have a non-zero value as q approaches
zero. The finite width at q=0 may indicate a significant plasmon
life-time decay mainly driven by enhanced Coulomb collision which
has never been expected for nearly free-electron gas.\cite{Nagao}
The importance of electron correlations for 1D plasmon decay
mechanism has recently been observed also for the 1D plasmon from
semi-metallic Au chains on the Si(111) surface.\cite{Nagao06}
However, details of the width dispersion exhibit different behavior
mainly due to the lack of the saturation of width beyond $\sim$0.05
\AA$^{-1}$ as observed from the Au chains,\cite{Nagao06} which might
originate from relative narrow range 0.004
\AA$^{-1}$$\leq$q$\leq$0.034 \AA$^{-1}$ of our measurements. Precise
measurements at low q range for the Au chains are necessary in order
to generalize the intriguing property of elementary excitations in
interacting quasi-1D electron systems. The delicate nature of 1D
plasmon, however, can appear different when the metallicity of 1D
systems are different, i.e., semi-metallic as for the Au
chains\cite{Himp} and metallic as for the In nanowires considered
here.

The damping of life time (or equivalently $\zeta^{-1} (q)$) due to thermal fluctuation has also been observed in the
single walled carbon nanotubes.\cite{Ish} Ishii {\it et al.} reported that single walled carbon nanotubes show the NFL
behavior enhanced with decreasing temperature. The finite spectral intensity at Fermi level at room temperature has
been interpreted as a result of the NFL features by thermal fluctuation smearing the pseudo gap of the NFL. The reduced
spectral intensity near Fermi level observed also for these In nanowires may also be ascribed partially to thermal
fluctuation, which may be a signature of the smeared pseudo gap of the NFL phase.\cite{Yeom02} Zwick {\it et al.}
provides another example of the NFL phase in the 1D organic conductor TTF-TCNQ where a pseudo gap growing with
increasing temperature and the much reduced spectral weight near Fermi level are observed. These are considered as
evidence of the NFL phase despite the absence of the spin-charge separation.\cite{Zwick} We thus conclude that the In
nanowires, despite their non-ideal quasi 1D nature and finite temperature, reveals the NFL property through the
peculiar behavior of the width dispersion of intrasubband plasmon.

\section{Conclusion}
We have measured the dispersion of an intrasubband plasmon of the quasi-1D In nanowires self-assembled on the
Si(111)-7$\times$7 surface at room temperature. We observe quite anisotropic dispersions along the directions parallel
and perpendicular to the nanowires demonstrating the 1D nature of the intrasubband plasmon. The non-dispersive EELS
peak appearing in both directions is ascribed to an interband transition between three parallel surface bands near
Fermi level. The energy dispersion $\omega(q)$ of the unique dispersive peak agrees quite well with predictions by the
NFL theory and also by the RPA. The peculiar non-analytic width dispersion of the plasmon, however, as predicted only
by the NFL theory strongly supports the NFL nature of the In nanowires even though spin-charge separation is not
significant. One may observe similar non-analytic width dispersion of low energy excitations indicating the NFL feature
in other real 1D conducting systems.

\acknowledgements This work was supported by the Korea Research
Foundation Grants(KRF-2004-041-C00131 and KRF-2003-041-C00098).


\begin{thebibliography}{}
\bibitem{Voit} Voit J 1995 {\it Rep. Prog. Phys.} {\bf 58} 977
\bibitem{chang96} Chang A M, Pfeiffer L N and West K W 1996 {\it Phys. Rev. Lett.} {\bf 77} 2538
\bibitem{Ish} Ishii H {\it et al.} 2003 {\it Nature} {\bf 426} 540
\bibitem{Sego} Segovia P, Purdie D, Hengsberger M and Baer Y 1999 {\it Nature} {\bf 402} 504
\bibitem{Lo} Losio R, Altmann K N and Himpsel F J 2000 {\it Phys. Rev. Lett.} {\bf 85} 808; 2001 {\it ibid.} {\bf 86} 4632
\bibitem{Yeom02} Yeom H W, Horikoshi K, Zhang H M, Ono K and Uhrberg R I G 2002 {\it Phys. Rev. B} {\bf 65} 241307(R); 1999 {\it Phys. Rev. Lett.} {\bf 82} 4898
\bibitem{Kim} Kim C, Matsuura A Y, Shen Z -X, Motoyama N, Eisaki H, Uchida S, Tohyama T and Maekawa S 1996 {\it Phys. Rev. Lett.} {\bf 77} 4054
\bibitem{Aus} Auslaender O M, Steinberg H, Yacoby A, Tserkovnyak Y, Halperin B I, Baldwin K W, Pfeiffer L N and West K W 2005 {\it Science} {\bf 308} 88
\bibitem{Zwick} Zwick F, J\'{e}rome D, Margaritondo G, Onellion M, Voit J and Grioni M 1998 {\it Phys. Rev. Lett.} {\bf 81} 2974
\bibitem{Hager} Hager J, Matzdorf R, He J, Jin R, Mandrus D, Cazalilla M A and Plummer E W 2005 {\it Phys. Rev. Lett.} {\bf 95} 186402, and references therein
\bibitem{Sam} Samokhin K V 1998 {\it J. Phys.:Condens. Matter} {\bf 10} L533
\bibitem{Cap} Capurro F, Polini M and Tosi M P 2003 {\it Physica B} {\bf 325} 287
\bibitem{Felde} vom Felde A, Spr\"{o}sser-Prou J and Fink J 1989 {\it Phys. Rev. B} {\bf 40} 10181
\bibitem{Nagao} Nagao T, Hildebrandt T, Henzler M and Hasegawa S 2001 {\it Phys. Rev. Lett.} {\bf 86} 5747
\bibitem{Birtsch} Bertsch G F, Esbensen H and Reed B W 1998 {\it Phys. Rev. B} {\bf 58} 14031
\bibitem{Goni91} G\~{o}ni A R, Pinczuk A, Weiner J S, Calleja J M, Dennis B S, Pfeiffer L N and West K W 1991 {\it Phys. Rev. Lett.} {\bf 67} 3298
\bibitem{Sarma96} Das Sarma S and Hwang E H 1996 {\it Phys. Rev. B} {\bf 54} 1936; 1985 {\it ibid.} {\bf 32} 1401; 1992
{\it ibid.} {\bf 45} 13713
\bibitem{Bor97} Borges A N, Le\~{a}o S A and Hip\'{o}lito O 1997 {\it Phys. Rev. B} {\bf 55} 4680
\bibitem{Kuz03} Kuzmenko I, Gredeskul S, Kikoin K and Avishai Y 2003 {\it Phys. Rev. B} {\bf 67} 115331
\bibitem{Bunk99} Bunk O, Falkenberg G, Zeysing J H, Lottermoser L, Johnson R L, Nielsen M, Berg-Rasmussen F, Baker J
and Feidenhans'l R 1999 {\it Phys. Rev. B} {\bf 59} 12228
\bibitem{Kumpf00} Kumpf C, Bunk O, Zeysing J H, Su Y, Nielsen M, Johnson R L, Feidenhans'l R and Bechgaard K 2000 {\it Phys. Rev. Lett.} {\bf 85} 4916
\bibitem{Ahn04} Ahn J R, Byun J H, Koh H, Rotenberg E, Kevan S D and Yeom H W 2004 {\it Phys. Rev. Lett.} {\bf 93} 106401
\bibitem{Sak} Sakamoto K, Ashima H, Yeom H W and Uchida W 2000 {\it Phys. Rev. B} {\bf 62} 9923
\bibitem{Cho05} Cho J H, Lee J Y and Kleinman L 2005 {\it Phys. Rev. B} {\bf 71} 081310(R)
\bibitem{Lee02} Lee S S, Ahn J R, Kim N D, Min J H, Hwang C G, Chung J W, Yeom H W, Ryjkov S V and Hasegawa S 2002 {\it Phys. Rev. Lett.} {\bf 88} 196401
\bibitem{Laiten} Laitenberger P and Palmer R E 1996 {\it Phys. Rev. Lett.} {\bf 76} 1952
\bibitem{Tan} Tanatar B 1995 {\it Phys. Rev. B} {\bf 51} 14410
\bibitem{Sturm} Sturm K and Oliveira L E 1981 {\it Phys. Rev. B} {\bf 24} 3054
\bibitem{Nagao06} Nagao T, Yaginuma S, Inaoka T and Sakurai T 2006 {\it Phys. Rev. Lett.} {\bf 97} 116802
\bibitem{Himp} Barke I, Zheng F, R\"{u}gheimer T K and Himpsel F J 2006 {\it Phys. Rev. Lett.} {\bf 97} 226405
\end{thebibliography}
\end{document}